\documentclass[runningheads,a4paper]{llncs}

\usepackage{dsfont} 
\usepackage{stmaryrd} 
\DeclareMathAlphabet{\mathpzc}{OT1}{pzc}{m}{it} 
\usepackage{color}
\usepackage{graphicx}
\usepackage{listings}
\usepackage[american,ngerman]{babel} 
\usepackage{multirow}

\usepackage{url}
\usepackage[utf8]{inputenc}
\renewcommand\thanks[1]{\renewcommand\thefootnote{\fnsymbol{footnote}}\footnote{
    #1}\noindent}

\title{
  Constraint solving for high-level WCET
  analysis\thanks{ \raggedright This work has been partially supported
    by the Austrian Science Fund (Fonds zur F{\"o}rderung der
    wissenschaftlichen Forschung) under contract No P18925-N13,\\
    ``Compiler Support for Timing Analysis'',
    \url{http://costa.tuwien.ac.at/}\\
    and the 7th EU R\&D Framework Programme under contract No 215068,\\
    ``Integrating European Timing Analysis Technology'' (ALL-TIMES).
  }
}

\author{Adrian Prantl\inst{1} \and Jens Knoop\inst{1} \and Markus
  Schordan\inst{2} \and Markus Triska\inst{1} }

\institute{
  Institut für Computersprachen,\\
  Vienna University of Technology, Vienna, Austria\\
  email: \{adrian,knoop,triska\}@complang.tuwien.ac.at
  \and University of Applied Sciences Technikum Wien, Vienna, Austria\\
  email: schordan@technikum-wien.at
} 

\date{\today}

\begin{document}

\selectlanguage{american}

\maketitle
\setcounter{page}{77}

\begin{abstract} 
  The safety of our day-to-day life depends crucially on the correct
  functioning of embedded software systems which control the
  functioning of more and more technical devices. Many of these
  software systems are time-critical. Hence, computations performed
  need not only to be correct, but must also be issued in a timely
  fashion. \emph{Worst case execution time $($WCET$)$} analysis is
  concerned with computing tight upper bounds for the execution time
  of a system in order to provide formal guarantees for the proper
  timing behaviour of a system. Central for this is to compute safe
  and tight bounds for loops and recursion depths. In this paper, we
  highlight the TuBound approach to this challenge at whose heart is a
  constraint logic based approach for loop analysis.
\end{abstract}

\selectlanguage{american}

\section{Motivation}
Embedded software systems are virtually ubiquitous today to control
the functioning of technical devices we routinely use and rely on in
our day-to-day life. Many of these systems are safety-critical. Think
of applications in the avionics and automotive field such as
\emph{fly-by-wire} or its foreseeable companion technology
\emph{drive-by-wire}, where there is no longer any mechanical linkage
between the pilot stick and the steering gear of an aircraft or the
steering wheel and the tires of a car. Applications like these
demonstrate that it is not only the comfort and convenience of our
day-to-day life but also its safety, which depends crucially on the
correct functioning of these systems. Many of these systems are also
time-critical. This means that calculations performed by such a system
need not only to be correct but also have to be issued in a timely
fashion. \emph{Worst case execution time $($WCET$)$} analysis is
concerned with providing formal guarantees for the proper timing
behaviour of a system by computing tight upper bounds for the
execution time of a system.

State-of-the-art WCET analysis tools rely on supporting analyses to
provide them with information on the execution behaviour of the
program such as loop bounds or maximum recursion depths. Typically,
both steps are performed on the binary code of the program. While this
is in fact mandatory for the WCET analysis in the narrow sense in
order to get an upper bound of the execution time of the code that is
actually executed, it is not for the supporting analyses. %
This is important to note because it is usually more difficult to
implement and perform the supporting analyses on the binary code of a
program, since type or control-flow information, which is readily
available in the source code, is not in the binary code; it also
imposes a particular hardship on the programmer, when demanded to
provide information manually as is occasionally necessary e.g.~because
of the undecidability of the involved analysis problems.

The TuBound approach~\cite{TuBoundWCET08}, which we pursue in the
CoSTA project~\cite{CostaWWW}, is to improve on this by lifting the
supporting analyses for WCET analysis to the source code level of a
program. The information computed on this level and annotated in the
code is then conjointly transformed throughout the compilation and
optimization of the program to the binary code level to make it
accessible to the WCET analysis component of our TuBound
tool. Currently, all optimizations are performed on the source code
level, too. The transformed and optimized code is then fed into a
specific WCET-aware~\cite{engblom98facilitating} variant of the Gnu-C
compiler~\cite{Kirner:PTHESIS2003}, which is tailored for preserving
the validity of code annotations it is provided with in the compiled
code. The binary code it generates is finally passed to a retargetable
WCET analysis component, which computes the desired upper bound of the
execution time of the program in the worst case. Currently, this is
the WCET analyzer {\sc CalcWCET}$_{167}$~\cite{CalcWCETWWW}.

The outcome of the recent participation of the TuBound tool in the
\emph{2008 WCET Tool Challenge}, which has been held as a part of the
\emph{8th International Workshop on Worst-Case Execution Time Analysis
  $($WCET'08$)$} shows the practicality and the power of the TuBound
approach and tool~\cite{Challenge08}.
Analyzing the results of the WCET Tool Challenge shows that the key to
the success of the TuBound tool is the generality and precision of
the constraint logic based approach for loop analysis to fully
automatically compute safe and tight loop bounds and flow constraints
for most of the benchmark programs subject to this
challenge~\cite{ChallengeWWW}. The usage of logic and constraint-logic
programming was fundamental to achieve this and to obtain a stable
prototype in short time and with moderate effort.

In this paper, which is an elaborated version of a recent oral
presentation~\cite{ResAn08} at the \emph{Workshop on Resource Analysis
  $($ResAn'08$)$}, we focus on the essence of our constraint-logic
based loop analysis and its implementation in Prolog. Below, we
present a summary of key components of the TuBound tool including our
constraint-logic solver, before presenting our constraint-logic based
loop analysis in detail.

\section{TuBound Architecture: Key Components}

\subsection{SATIrE, ROSE and PAG}
Most important to implementing the source-to-source analysis and
optimization approach of TuBound are the usage and integration of the
SATIrE, LLNL-ROSE, and PAG systems of TU Vienna, Lawrence Livermore
National Laboratory (LLNL) and AbsInt GmbH, 
respectively~\cite{SchordanBH07,RoseWWW,Martin98}. SATIrE is the
\emph{Static Analysis Integration Engine}~\cite{SchordanDagstuhl:08},
which seamlessly connects the C++ source-to-source compiler
infrastructure LLNL-ROSE~\cite{SchordanQ03} with the \emph{Program
  Analysis Generator} PAG~\cite{Martin98}. In TuBound, this enables us
to create data-flow analyses which operate on the abstract syntax tree
(AST) of C++ programs. Moreover, SATIrE supports the import and export
of an external term representation of the AST using Prolog
syntax. This term representation is automatically annotated with any
results from preceding analysis steps and contains all necessary
information to correctly unparse the program, including line and
column information for each expression. This term representation is
the key to specifying analyses and program transformations in Prolog,
which we make strong use of in TuBound. We are thereby benefiting from
many advantages over using C++ for the specification, including
pattern matching and access to tools and methods offered by the world
of logic programming. Most outstandingly, in TuBound this has been
used to implement a flow constraint analysis by means of our
generalized finite domain constraint solver
(cf.~Section~\ref{sec:CLP}), a loop bound analysis written in
SWI-Prolog (cf.~Appendix), and an interprocedural interval analysis
specified with PAG~\cite{Martin98} (cf.~Appendix).

\subsection{CLP(FD)}
\label{sec:CLP}
Generally speaking, \emph{constraint logic programming over finite
  domains}, denoted as CLP(FD), is a declarative formalism for
modeling and solving combinatorial problems over integers. A
\emph{constraint satisfaction problem $($CSP$)$} consists of:

\begin{itemize}
\item a set $X$ of variables, $X = \{x_1, \dots, x_n\}$

\item for each variable $x_i$, a set $D(x_i)$ of values that $x_i$ can
 assume, which is called the \textit{domain} of~$x_i$

\item a set of \textit{constraints}, which are simply relations among
 variables in~$X$, and which can further restrict their domains.
\end{itemize}

In TuBound, we use the solver library \emph{clpfd}~\cite{clpfd:wlp2008}. This
is a generalised CLP(FD) solver that we developed, and which was
recently included in the SWI-Prolog
distribution~\cite{Wielemaker:03b}. Two new features which are
implemented in this solver make it especially well-suited for the loop
analysis presented in Section~\ref{sec:clp-analysis}: First, the
solver can reason over arbitrarily large integers, and can thus also
be used to analyse a large number of nested loops that can range over
large bounds. Second, constraint propagation in our solver always
terminates. While this weakens propagation when domains are still
unbounded, this property guarantees that the loop analysis itself
always terminates. It is this property, which makes the \emph{clpfd}
solver particularly useful for the TuBound approach described next.

\section{Loop constraint analysis in TuBound}

\subsection{Preliminaries}

TuBound derives loop bounds and constraints for
\emph{iteration-variable based loops}. This type of loop is very
common in embedded applications. The
\emph{debie}~program~\cite{Challenge08}, a real-world space-craft
control system used in the WCET tool challenge, for example, contains
88\% iteration-variable based loops. 
We call a loop $L$ iteration-variable based if
\begin{itemize}
\item it is preceded by an initialization statement $[i := a]^{l_1}$,
\item it contains at least one exit condition $[i\ rel\
  b]^{l_2}$,
\item it contains exactly one monotone iteration step statement 
      $[i := i + c]^{l_3}$,
\end{itemize}
where $i$ is an integer variable that is not a field member with a
scope larger than the loop body ($L \subseteq scope(i)$). The labels
$l_1,l_2,l_3$ are used to reference these statements.

We assume that the program has been run through an interval
analysis\footnote{This analysis is implemented in TuBound, too, but it
  is beyond the core focus of this paper. Details can be found in the
  Appendix.}  that generates variable-interval pairs for each
variable $v$ at each sequence point, denoted $v_{\{min,max\}}$. In
order to classify the results, we call the value of a variable at a
given location \emph{loop invariant}, if the value does not change
over all paths through $L$.  We call the value of a variable
\emph{constant}, if $v_{min} = v_{max}$. This implies that every
constant variable is also loop invariant.
Having this information, we can verify that the loop satisfies the
following safety conditions:
\begin{itemize}
\item[$C1$.] There is no statement $s \in L\setminus\{l_1,l_2,l_3\}$ where $i$
  appears on the left-hand side.
\item[$C2$.] There is no statement $s \in scope(i)$ in that the
  address of $i$ is taken.
\item[$C3$.] The loop must not be an infinite loop, i.e.~the
  analyzable intervals of $a$ and $b$ must either be disjoint or
  overlapping in at most one value. Further, the \emph{direction} of the loop
  must be unambiguous $(sgn(c_{min}) = sgn(c_{max}))$ and the property
  $sgn(b-a) = sgn(c_{min})$ must hold.
\item[$C4$.] The exit condition's relational operator $rel$ must induce
  a partial order $\{\leq,\geq\}$. For equivalence operators
  $\{=,\neq\}$ it is also necessary to prove that the loop terminates
  at all, before they can safely be replaced with $\{\leq,\geq\}$. For
  the case that $a, b, c$ are \emph{loop invariant} and
  \emph{constant},
  $$b-a\pmod{c} \equiv 0$$
  is a sufficient condition for termination. The operators $<$ and $>$ can be
  transformed by adding $\pm 1$ to $b$.

\end{itemize}

\subsection{The constraint analysis for nested loops}
\label{sec:clp-analysis}

TuBound contains an implementation of a loop-bound algorithm that
works for nested loops. If the iteration space described by the
iteration variables is rectangular or cuboid-shaped, the resulting
bounds will even be optimal. Often, however, the iteration variables
of nested loops depend on each other, forming e.g.~a triangular
iteration space. Loop bounds would then be an overestimation of the
iteration space, describing the enclosing rectangle. It is thus
desirable to formulate more general \emph{flow constraints} in
addition to loop bounds.  The flow constraints we are generating
describe the execution counts of the loop bodies in relation to the
scope containing the outermost loop.
Our constraint analysis works by transforming the whole loop nest into
finite domain logic constraints.

Each loop in the loop nest must be iteration-variable based. In
contrast to the traditional loop bound analysis, a few additional
restrictions are imposed on the loop: The step size must be
loop invariant. When the loop has a stride greater than $1$ ($|c|>1$), $a$
should be constant. Otherwise the results produced by the analysis
will be an overestimation which is bounded by a factor of
$a_{max}-a_{min}$.  Furthermore, the exit test expression must either
test for $\leq$ or $\geq$; $a < b$ can be transformed into the
equivalent $a \leq b - 1$.

The algorithm works recursively, beginning with the outermost loop.
First, a new logic variable \verb|I| is created that is associated
with the iteration variable. Then, the $init$, $test$ and $step$
statements are translated into constraints, as sketched
in Table~\ref{tab:constraints}.
\begin{table}
  \centering
  \begin{tabular}{llll}
    Direction\hspace{.5em} & Init                  & Test & Step \\ \hline
    up & \verb|I #>= InitExpr|\hspace{1em} &\verb|I #=< TestExpr|\hspace{.5em} 
      &\multirow{2}{*}{ {\tt (I-InitExpr) mod StepExpr \#= 0} } \\
    down      & \verb|I #=< InitExpr| & \verb|I #>= TestExpr| & \\
    \hline
  \end{tabular}
  \vspace{1em}
  \caption{Deriving the constraints}
  \label{tab:constraints}
\end{table}
The remaining arithmetic expressions can then recursively be
translated into corresponding constraints. After the constraints are
posted, the constraint solver is used to report the number $n$ of
possible combinations of all iteration variables that were encountered
so far. 
Since explicit enumeration of all solutions can be infeasible, we added a
new labeling option \emph{upto\_in} to our constraint solver, which can
be used to count the number of possible instantiations if all
remaining constraints are trivial. With this method, the running time
and memory consumption of the solver is no longer depending on the
size of the iteration space.

The resulting $n$ is then an upper bound for the number of
times the current (=innermost regarded) loop is executed relative to
the scope containing the outermost loop.  If the constraint analysis
is applied to a single loop only, the resulting constraint degenerates
into a loop bound.

By using this approach, we can leverage a great deal of features from
our constraint solver for the loop analysis:
\begin{itemize}
\item The order in which the constraints are posted does not influence
  the behaviour of the solver.
\item The termination of the constraint solver is guaranteed.
\item The strategy of the solver can be customized through labeling
  options to improve its efficiency (cf.~Section~\ref{sec:example}).
\item Through the implicit enumeration of the iteration space the
  results are generally more precise than those of the traditional loop
  bound analysis.
\end{itemize}
Since much of the complexity is offloaded into the constraint solver,
the implementation is very concise and easy to maintain.

\subsection{Example}
\label{sec:example}

We illustrate the general principle using the following loop nest, for
which we want to determine the number of times the inner loop is
executed:
\begin{verbatim}
  for (i = 0; i < 10; ++i)
    for (j = i; j > 0; j -= 2)
\end{verbatim}
By translating the loop nest accordingly, we get the following
constraint program:
\begin{verbatim}
I #>= 0, I #< 10, I mod 1 #= 0,
J #=< I, J #> 0, (J-I) mod 2 #= 0,
findall((I,J),labeling([], [I,J]), IS),
length(IS, IterationCount).
\end{verbatim}
By solving the constraint system, we explicitly enumerate the iteration
space \verb|IS| described by $(i,j)$:
\begin{verbatim}
[ (1, 1), 
  (2, 2), 
  (3, 1), (3, 3), 
  (4, 2), (4, 4), 
  (5, 1), (5, 3), (5, 5), 
  (6, 2), (6, 4), (6, 6), 
  (7, 1), (7, 3), (7, 5), (7, 7), 
  (8, 2), (8, 4), (8, 6), (8, 8), 
  (9, 1), (9, 3), (9, 5), (9, 7), (9, 9) ]
\end{verbatim}
The number of pairs in the iteration space is then an upper bound for
the innermost loop body. In our case, exactly 25 times.
For larger bounds, explicit enumeration of all solutions is
infeasible. We therefore added a new labeling option to our constraint
solver, which can be used to count the number of possible
instantiations if all remaining constraints are trivial. Thus we can
reduce or avoid explicit enumeration in many cases. For example:
\begin{verbatim}
I #>= 0, I #=< 10000, 
J #>= 0, J #=< 500,
labeling([upto_in(IterationCount)], [I,J]).
\end{verbatim}
yields~$\textrm{IterationCount} = 5010501$.

\section{Experimental results}

To evaluate the constraint analysis, we compare its power with that of
a pure loop bound analysis. This second analysis was implemented at an
earlier stage and is also part of TuBound (cf.~Appendix). The loop
bound analysis works by solving linear equations that are derived from
the loop parameters.

\begin{table}
  \centering
  \begin{tabular}{lrrrrr}
    \hline
    Benchmark & Loops & \hspace{2em}Loopbounds & \hspace{1em}Runtime & \hspace{2em}Constraints & \hspace{1em}Runtime \\
    \hline
    adpcm & $18$ & $83.3\%$ & $0.02s$ & $83.3\%$ & $0.02s$ \\
    bs & $1$ & $0\%$ & $<0.01s$ & $0\%$ & $<0.01s$ \\
    bsort100 & $3$ & $100.0\%$ & $<0.01s$ & $100.0\%$ & $<0.01s$ \\
    cnt & $4$ & $100.0\%$ & $<0.01s$ & $100.0\%$ & $<0.01s$ \\
    compress & $7$ & $28.5\%$ & $0.01s$ & $14.2\%$ & $0.08s$ \\
    cover & $3$ & $100.0\%$ & $0.01s$ & $100.0\%$ & $0.01s$ \\
    crc & $3$ & $100.0\%$ & $<0.01s$ & $100.0\%$ & $<0.01s$ \\
    des & $10$ & $90.0\%$ & $0.09s$ & $90.0\%$ & $0.09s$ \\
    duff & $2$ & $50.0\%$ & $<0.01s$ & $50.0\%$ & $<0.01s$ \\
    edn & $12$ & $100.0\%$ & $0.02s$ & $91.6\%$ & $0.05s$ \\
    expint & $3$ & $100.0\%$ & $<0.01s$ & $100.0\%$ & $<0.01s$ \\
    fdct & $2$ & $100.0\%$ & $0.01s$ & $100.0\%$ & $0.01s$ \\
    fft1 & $11$ & $54.5\%$ & $<0.01s$ & $18.1\%$ & $0.41s$ \\
    fibcall & $1$ & $100.0\%$ & $<0.01s$ & $100.0\%$ & $<0.01s$ \\
    fir & $2$ & $50.0\%$ & $0.03s$ & $50.0\%$ & $0.03s$ \\
    insertsort & $2$ & $50.0\%$ & $<0.01s$ & $0\%$ & $<0.01s$ \\
    janne\_complex & $2$ & $50.0\%$ & $<0.01s$ & $0\%$ & $<0.01s$ \\
    jfdctint & $3$ & $100.0\%$ & $<0.01s$ & $100.0\%$ & $<0.01s$ \\
    lcdnum & $1$ & $100.0\%$ & $<0.01s$ & $100.0\%$ & $<0.01s$ \\
    lms & $10$ & $60.0\%$ & $0.02s$ & $60.0\%$ & $0.01s$ \\
    ludcmp & $11$ & $100.0\%$ & $0.01s$ & $81.8\%$ & $0.01s$ \\
    matmult & $5$ & $100.0\%$ & $<0.01s$ & $100.0\%$ & $<0.01s$ \\
    minver & $17$ & $94.1\%$ & $0.01s$ & $82.3\%$ & $0.28s$ \\
    ndes & $12$ & $100.0\%$ & $0.04s$ & $100.0\%$ & $0.04s$ \\
    ns & $4$ & $100.0\%$ & $0.02s$ & $100.0\%$ & $0.02s$ \\
    nsichneu & $1$ & $0\%$ & $0.06s$ & $0\%$ & $0.06s$ \\
    qsort-exam & $6$ & $0\%$ & $<0.01s$ & $0\%$ & $<0.01s$ \\
    qurt & $1$ & $100.0\%$ & $<0.01s$ & $100.0\%$ & $<0.01s$ \\
    recursion & $0$ & -- & $<0.01s$ & -- & $<0.01s$ \\
    select & $4$ & $0\%$ & $<0.01s$ & $0\%$ & $<0.01s$ \\
    statemate & $1$ & $0\%$ & $0.02s$ & $0\%$ & $0.03s$ \\
    sqrt & $1$ & $100.0\%$ & $<0.01s$ & $100.0\%$ & $<0.01s$ \\
    st & $5$ & $100.0\%$ & $<0.01s$ & $100.0\%$ & $<0.01s$ \\
    whet & $10$ & $100.0\%$ & $0.01s$ & $80.0\%$ & $0.02s$ \\
    \hline
    Total Percentage & & $80.8\%$ & & $72.3\%$ & \\
    \hline
  \end{tabular}
  \vspace{1em}
  \caption{Results for the Mälardalen WCET benchmark suite}
  \label{tab:benchmark}
\end{table}

\begin{table}
  \centering
  \begin{tabular}{lrrrrr}
    \hline
    Benchmark & Loops & \hspace{2em}Loopbounds & \hspace{1em}Runtime & \hspace{2em}Constraints & \hspace{1em}Runtime \\
    \hline
    class & $2$ & $100.0\%$ & $0.01s$ & $50.0\%$ & $0.01s$ \\
    hw\_if & $3$ & $100.0\%$ & $0.01s$ & $33.3\%$ & $0.33s$ \\
    classtab & $0$ & -- & $0.01s$ & -- & $0.01s$ \\
    measure & $14$ & $85.7\%$ & $0.02s$ & $71.4\%$ & $0.02s$ \\
    debie & $1$ & $0\%$ & $<0.01s$ & $0\%$ & $<0.01s$ \\
    tc\_hand & $13$ & $92.3\%$ & $0.03s$ & $92.3\%$ & $0.03s$ \\
    harness & $43$ & $76.7\%$ & $0.08s$ & $76.7\%$ & $0.08s$ \\
    telem & $6$ & $100.0\%$ & $0.01s$ & $66.6\%$ & $0.31s$ \\
    health & $11$ & $81.8\%$ & $0.03s$ & $63.6\%$ & $2.30s$ \\
    \hline
    Total Percentage & & $82.6\%$ & & $72.1\%$ & \\
    \hline
  \end{tabular}
  \vspace{1em}
  \caption{Results for the debie WCET benchmark suite}
  \label{tab:debie}
\end{table}

We are using the standardized WCET benchmark suite from Mälardalen
University~\cite{BenchmarksWWW}, consisting of over 30 prototypical
embedded programs and the \emph{debie} benchmark from the WCET Tool
Challenge 2008~\cite{Challenge08}. Since the prerequisites for
applying the constraint analysis are slightly more restrictive than
for the loop bound analysis, we expect the constrained loops to be a
subset of the bounded loops. This is confirmed by the results, which
are shown in Table~\ref{tab:benchmark} for the Mälardalen University
benchmarks and in Table~\ref{tab:debie} for the \emph{debie}
benchmark.

The first column lists the name of the benchmark, the second column
the number of loops that are contained in that benchmark. Column three
gives the percentage of loops that could be analyzed with the
traditional loop bound algorithm discussed in the Appendix. The
running times of the algorithm in seconds\footnote{Measurements were
  made on a 3 GHz Xeon, running SWI-Prolog 5.6.59 under Linux.}  is
shown in the next column. The last two columns contain the percentage
of loops that could be analyzed with the constraint analysis and the
corresponding running times. From examining the table we can see that
the constraint analysis can analyze almost 90\% of the loops that are
analyzable with the traditional approach and more than 70\% of all
loops contained in the benchmarks.  Moreover, the constraint analysis
inherently outperforms the traditional approach on nested loops with
non-rectangular iteration space, due to the higher expressivity of
flow constraints.

When comparing the runtime performance of the two approaches, it is
apparent that the loop bound analysis mostly depends on the depth of
the $init$, $test$ and $step$ expressions, whereas the worst-case
running time of the constraint analysis is correlated with the size of
the iteration space, if the solver has to fallback to enumeration.
For typical embedded code that we target with TuBound, this has little
significance, since analyzing even the outliers is a matter of
seconds. The average execution time of both analyses together is well
below one second on current hardware. Methods like
in~\cite{AlbertAGP08} could be used to complement this with a more
theoretical performance statement.

\section{Conclusion and  perspectives}

We have presented our design and implementation for a generalized loop
constraint analysis, which plays an important role as a supporting
analysis in our WCET analysis tool, TuBound. Our results demonstrate
that this analysis can determine tighter flow constraints for nested
loops than our traditional loop bound analysis.

Loop bound and constraint analysis together succeed in analyzing both
standardized benchmarks and real-world programs such as the debie
spacecraft control system used in the WCET Tool Challenge
2008~\cite{Challenge08}, with only a handful of necessary manual
annotations remaining.

Since the constraint analysis can also be adopted to derive loop
bounds, we plan to replace the traditional loop bound analysis
implementation by the constraint analysis eventually.  Moreover, by
offloading complexity into a separately maintained library, the
analysis will automatically benefit from future improvements made to
the solver. Thanks to its clean interface, we also retain the
possibility to switch to different constraint solvers in the future.



\clearpage
\appendix
\begin{center}
  {\bf APPENDIX}
\end{center}

\subsection*{Interval Analysis}
The interval analysis is an interprocedural data-flow problem. The
variant we implemented in TuBound is an extension of the constant
propagation analysis specified by Nielson, Nielson and
Hankin~\cite{NNH99}. Earlier work on interval analysis, also called
value range propagation, was done by Harrison~\cite{Harrison77} and
also by Cousot and Cousot~\cite{Cousot77}.  The design parameters are
sketched in Table~\ref{tab:interval}. Just as constant propagation,
the interval analysis is a forward-directed data-flow problem. The
carrier of the analysis is a lattice of pairs of integers that are
mapped to each integer variable. The members of the pairs denote the
lower and upper bounds of the variables, respectively. If a bound is
unknown, it is reported as $\pm\infty$. The $\bot$ element of the
lattice means that a value has not yet been calculated, whereas $\top$
represents an unknown bound, which is equivalent to $(-\infty,
\infty)$. At a control-flow join, the combine function is applied
pairwise for each variable and merges the interval information coming
from the different branches. The transfer functions for each statement
capture the ramifications of the statement on the $State$ lattice by
abstractly interpreting the statement with interval arithmetic
(function $\mathcal{A}_{Itvl}$)~\cite{Alefeld74}. The widening
operator, which is used to speed up the fixed-point search is defined
very aggressive, and can be used to fine-tune the trade-off between
execution speed and analysis precision.

The transfer functions for conditional branches return different
results for the \emph{true} and \emph{false} edges. If the branch
condition statically evaluates to either $(1,1)$ or $(0,0)$, the state
for the other branch is set to $\bot$, such that dead code can not
influence the analysis result for live branches.

\newcommand{\Aint}[1]{\mathcal{A}_{Itvl}\llbracket #1\rrbracket\sigma}

\begin{table*}
  \begin{tabular}{ll}
    \hline
    Direction: & $forward$ \\
    Lattice: & $State = (Var \rightarrow (\mathds{Z}^{-\infty},\mathds{Z}^\infty),
      \sqsubseteq, \sqcup, \sqcap, \bot, \lambda x.(-\infty, \infty))$ \\
    Init function: & $\lambda x.(-\infty, \infty)$ \\
    Combine function: & $comb((a_{min}, a_{max}), (b_{min}, b_{max})) = 
    (min(a_{min}, b_{min}), max(a_{max}, b_{max}))$\\
    Widening operator: & $widen((a_{min}, a_{max}), (b_{min}, b_{max})) = 
    (c_{min}, c_{max})$ \\
    & where $c_{min} = \left\{
      \begin{array}{ll}
        a_{min} & $if$\ a_{min} = b_{min} \\
        -\infty & $otherwise$ \\
      \end{array}
      \right.$ \\
    & \phantom{where} $c_{max} = \left\{
      \begin{array}{ll}
        a_{max} & $if$\ a_{max} = b_{max} \\
        \infty & $otherwise$ \\
      \end{array}
      \right.$ \\

    &\\
    Transfer functions: & 
    $[x := a]^l : f^{Itvl}_l(\sigma) = \left\{
    \begin{array}{ll}
      \bot & $if$\ \sigma = \bot \\
      \sigma [x\mapsto \Aint{a}] & $otherwise$ \\
    \end{array}
    \right.$ \\
    &\\
    & $[$if$ (c)]^l_{edge} : f^{Itvl}_l(\sigma) = \left\{
    \begin{array}{ll}
      \bot & $if$\ \sigma = \bot \\
      \bot & $if$\ [\Aint{a}] = edge \\
      f^{Itvl}_{l'}(\sigma), [c]^{l'} & $otherwise$ \\
    \end{array}
    \right.$ \\
    & where \\
    & $\begin{array}{lll}
      \Aint{x} &=& \sigma(x) \\
      \Aint{n} &=& (n, n) \\
      \Aint{a\ op\ b} &=& \Aint{a}\ op_{Itvl}\ \Aint{b} \\
    \end{array}$ \\
   &\\
   Interval arithmetic: & \multirow{2}{*}{
   $\begin{array}{rlll}
     +_{Itvl}(a, b) &=& 
     (a_{min}+b_{min}, a_{max}+b_{max})\\
     -_{Itvl}(a, b) &=& 
     (a_{min}-b_{max}, a_{max}-b_{min})\\
     \cdot_{Itvl}(a, b) &=& 
       (min(a_{min}\cdot b_{min}, a_{min}\cdot b_{max}), 
        max(a_{max}\cdot b_{min}, a_{max}\cdot b_{max}))\\
     /_{Itvl}(a, b) &=& 
     (min(a_{min}/b_{min}, a_{min}/b_{max}), max(a_{max}/b_{min}, a_{max}/b_{max}))\\
     =_{Itvl}(a, b) &=& \left\{
       \begin{array}{rl}
         a_{min} = b_{min} & $if$\ a_{min} = a_{max} \wedge b_{min} = b_{max} \\
                   false & $if$\ a_{max} < b_{min} \vee   a_{min} > b_{max} \\
                    \top & $otherwise$\\          
       \end{array}
     \right.\\
     \neq_{Itvl}(a, b) &=& \left\{
       \begin{array}{rl}
         a_{min} \neq b_{min} & $if$\ a_{min} = a_{max} \wedge b_{min} = b_{max} \\
                      true  & $if$\ a_{max} < b_{min} \vee   a_{min} > b_{max} \\
                       \top & $otherwise$\\          
       \end{array}
    \right. \\
     <_{Itvl}(a, b) &=& \left\{
       \begin{array}{rl}
         true  & $if$\ a_{max} < b_{min} \\
         false & $if$\ a_{min} \geq b_{max} \\
         \top  & $otherwise$\\          
       \end{array}
    \right. \\
     ... &&\\
   \end{array}$ } \\
   \vspace{15em}&\\
    \hline
  \end{tabular}
  \vspace{1em}
  \caption{Sketched specification of the interval analysis}
  \label{tab:interval}
\end{table*}

The accuracy of the interval analysis can further be improved by
increasing the memory and run-time budget: It can be modified to
report a set of possible intervals instead of one merged interval for
each variable.

\subsection*{Traditional loop bound analysis}

The loop bound analysis is a control flow insensitive analysis that
builds upon the results of the above interval analysis. The analysis
takes as input
\begin{enumerate}
\item an iteration-variable based loop $L$,
\item variable intervals
\item and context information (such as the scope of $i$).
\end{enumerate}
The analysis works on all iteration-variable based loops, with the
restriction that the step size must be either positive or negative:
$$sgn(step_{min}) = sgn(step_{max})$$
The result of the analysis is an upper bound $n$ for the number of
times the loop entry is executed in relation to its direct predecessor
statements outside of the loop, where $excnt(s)$ denotes the execution
count of statement $s$:
$$ \sum_{p \in pred(c) \setminus L}{excnt(p)} \leq n * excnt(c)$$
Since the discrete function described by the iteration step statement
is monotone and its gradient is constant, we can set up the following
equation for the loop bound:
$$n = \frac{val_{max} - val_{min}}{|val_{stepsize}|}$$
where $val_{min}$, $val_{max}$ are lower and upper bounds for $i$,
whereas $val_{step}$ is the minimum step size of $i$ on a path through
the loop $L$. We call these values \emph{loop parameters}.
To derive the loop parameters, it is necessary to examine the
relational operator of the exit condition, which must be one of
$<,>,\leq,\geq$.

\begin{table} 
  \centering
  \begin{tabular}{ll}
    \hline
    case $rel$ of & \\
    &$\begin{array}{lll}
      <:   & LowExpr = a, & HighExpr = b\\
      \leq:& LowExpr = a, & HighExpr = b+1\\
      >:   & LowExpr = b, & HighExpr = a\\
      \geq:& LowExpr = b, & HighExpr = a-1\\
    \end{array}$\\
    \\
    $StepExpr = c$\\
    \hline
  \end{tabular}
  \vspace{1em}
  \caption{Deriving the loop parameters}
  \label{tab:parameters}
\end{table}

As shown in Table~\ref{tab:parameters}, the assignment of $LowExpr$
and $HighExpr$ depends on the direction of the loop. In our
implementation, concrete values of the loop parameters are calculated
in two phases:

\begin{enumerate}
\item \emph{Simplify.} In this phase, algebraic identities are
  exploited to simplify the expression
  ${(HighExpr-LowExpr)}/{StepExpr}$. This is implemented by a set of
  rewrite rules that are applied to the expression until a fixed point
  is reached. This simplification operates on purely symbolic
  expressions and disregards the analyzed intervals of variables. It
  can, however, use the information that an expression is
  \emph{loop invariant} or \emph{constant}, i.e.~no variable occurring in
  it appears on the left-hand side of any statement in $L$.
\item \emph{Evaluate.} Using the results of the interval analysis as
  state, we can evaluate the simplified expression using interval
  arithmetic~\cite{Alefeld74} ($\mathcal{A}_{Itvl}$). The return value
  is an interval $(m,n)$ where $n$ is the upper bound for the
  iteration count of the loop $L$.
\end{enumerate}

The complexity of this algorithm is bounded by the number of exit
conditions in the loop, the depth of $Low$, $High$ and $Step$
expressions and the number of rules in the simplification term
replacing system.

\end{document}